# Gold nanodoughnut as an outstanding nanoheater for photothermal applications

Javier González-Colsa[1], Guillermo Serrera[1], José María Saiz[1], Dolores Ortiz[1], Francisco González[1], Fernando Bresme[2], Fernando Moreno[1] and Pablo Albella[1*]

[1]*Group of optics, Department of Applied Physics, University of Cantabria, 39005, Santander, Spain.*
[2]*Department of Chemistry, Molecular Sciences Research Hub, Imperial College London, W120BZ, London, United Kingdom.*

*pablo.albella@unican.es

**Abstract:** Photoinduced hyperthermia is a cancer therapy technique that induces death to cancerous cells via heat generated by plasmonic nanoparticles. While previous studies have shown that some nanoparticles can be effective at killing cancer cells under certain conditions, there is still a necessity (or the need) to improve its heating efficiency. In this work, we perform a detailed thermoplasmonic study comparing the most effective nanoparticle geometries up to now with a doughnut-shaped nanoparticle, demonstrating that the latter exhibits a superior tunable photothermal response in practical illumination conditions, i.e., unpolarized light. Furthermore, we show that nanoparticle heating in fluidic environments, i.e., nanoparticles undergoing Brownian rotations, strongly depends on the particle orientation with respect to the illumination source. We conclude that the heating performance of nanodoughnuts is outstanding, with a temperature increment 35% higher than the second best nanoheater (nanodisks). Furthermore, nanodoughnuts feature a weak dependence on orientation, being therefore, ideal candidates for photothermal therapy applications. Finally, we present a designing guide, covering a wide range of toroid designs, which can help on its experimental implementation.

## 1. Introduction

Cancer diseases constitute a major threat to public health, being one of the main causes of death worldwide. Around 18 million of new cancer cases and 9.6 million related deaths were reported in 2018 [1,2], with a rapidly growing trend in recent times [3,4]. While current medical treatments focus on surgical resection, chemotherapy, and radiotherapy; these methods are not sufficient for a complete cure [5]. Thus, novel methods with higher effectivity and side effect free are sought.

Among the existing alternative therapies, hyperthermia treatments involve local heating at 42-48ºC [6], leading to inhibition of protein activation and DNA synthesis [7,8] inducing cell death, with widely spread approaches like radiofrequency or microwave ablation [9]. However, macroscopic heating methods might affect large tissue areas, resulting potentially in the damage of healthy tissue as well as other side effects [10]. On the other hand, Photothermal Therapy (PTT), using absorbing nanoparticles as microscopic photothermal agents, can focus high temperatures in the region around the cancerous cells, hence preventing the damage of healthy tissue and improving the therapeutic effectivity of heating [11].

The success of the PTT relies strongly on the heating source, i.e., the optically absorbing agent employed. A wide range of materials, including bovine serum albumin heterojunctions [12], graphene nanoparticles and carbon nanotubes [13], are being investigated for PTT applications, but the most widespread and promising PTT agents are based on Localized Surface Plasmon (LSP) resonances [14]. These nanoparticles allow for local heat generation by resistive losses enhancement caused by light absorption at the plasma eigenfrequency. The plasma eigenfrequency of a particle depends strongly on its composition, shape and size. All these variables can be exploited to produce highly tunable heating agents. Most noble metals exhibit LSP resonances in the UV-Vis-NIR range [15], but gold is the preferred material, due to its biocompatibility, weak cytotoxicity, and low reactivity. Furthermore, gold LSP resonances can be taken to the NIR, close to the so called "biological windows" (700-900 nm, 1000-1400 nm) [16], in which human tissues are dispersive – causing light depolarization– but low absorbing. Therefore, the design of particles with main absorption in this wavelength range is a key objective in PTTs, as they provide optimum heating conditions for the nanoparticles minimizing the non-selective heating of healthy tissues.

LSP resonances of nanoparticles can be controlled both, in magnitude and spectral position, by changing its shape and size. These variables therefore are key to optimize the PTT performance[17–19]. The rational design of nanoparticles drives many theoretical [20] and experimental efforts[21]. Among the most popular geometries[19], nanorods [22–24], core-shell nanoparticles [25–27] and nanostars [28,29], gold nanorods have attracted significant attention since they can be easily synthesized, producing monodisperse particles with well-defined aspect ratios and sizes below 120 nm [30–33], as nanoparticles should be small enough to penetrate small capillaries and get adsorbed to cells. Spherical gold nanoparticles feature resonances closer to the visible spectrum, but they can be combined into clusters [34,35] that lead to a redshift, and resonances closer to those corresponding to the biological windows. The collective effects can also enhance the performance of isolated nanoparticle thermo-optical responses giving rise to bigger temperatures as reported in [36,37]. Another aspect that has been covered with this type of systems is the impact of nanoscale curvature, which is relevant in small nanoparticles, on the efficiency of heat transfer from the particle to the solution. Recently, it was shown that nanoscale curvature influences the interfacial thermal conductance, resulting in enhanced thermal transport from particle to fluids [38].

Anisotropic structures provide an ideal approach to tune the spectral thermal response of nanoparticles [39]. However, when these particles diffuse freely in solution, their absorption is strongly dependent on the nanoparticle orientation with respect to the direction of the light source [39,40], affecting significantly their heating performance. Toroidal nanoparticles, whose fabrication in colloid has been recently demonstrated [24,41], have also been explored [42] compared with rods and rings.

In this work, we demonstrate that doughnut shaped nanoparticles, as single heaters, outperform the most used PTT agents, offering high spectral tunability, high performance of light into heat conversion and weak dependence of heating on orientation in solution. With this purpose we perform an exhaustive comparison of the spectral photothermal response of the most relevant gold nanostructures with a doughnut-shaped nanoparticle. Finally, a guide to design nanodoughnuts with tunable heating properties is provided.

## 2. Theoretical background

One of the most efficient ways to understand photothermal effects in plasmonic nanoparticles was introduced by Govorov et al. [43] The heat equation for a single spherical nanoparticle was solved and the steady state temperature as a function of the radial distance r can be derived analytically, with the well-known expression

$$\Delta T(r) = \frac{V_{NP} Q}{4\pi k_0 r} \tag{1}$$

where $V_{NP}$ is the nanoparticle volume, $r > R_{NP}$ is the radial distance, $R_{NP}$ is the nanoparticle radius and $k_0$ is the thermal conductivity of the surrounding medium. Here, the heat losses $Q$ in the particle can be obtained by means of the Joule effect expression

$$Q(\vec{r}) = \langle \vec{J}(\vec{r},t) \cdot \vec{E}(\vec{r},t) \rangle \tag{2}$$

Where $\vec{E}$ is the electric field inside the sphere, $\vec{J}$ is the current density, $\vec{r}$ is the position vector, and t is time. In most cases, intricate nanoparticle geometries prevent an analytical solution of the scattering problem in the nanoparticle. However, for a spherical nanoparticle, the electric field inside the sphere can be obtained by means of the quasi-static approximation [44], where the sphere size is much smaller than the excitation wavelength. Under this approximation, the $Q$ value can be obtained as

$$Q = \frac{\omega}{8\pi} E_0^2 \left| \frac{3\epsilon_0}{2\epsilon_0 + \epsilon_{NP}} \right|^2 \mathfrak{Im}[\epsilon_{NP}] \tag{3}$$

where $\omega$ is the angular frequency, $E_0$ the electric field amplitude, $\epsilon_0$ is the surrounding medium permittivity and $\epsilon_{NP}$ the permittivity of the nanoparticle. This model is useful in the study of microfluidic colloidal particles, as particles require sizes below 120 nm and the biological windows wavelengths are about 6-10 times this size.

## 3. Results and discussion

### 3.1 Geometry comparison

Most experiments aimed at photoinduced hyperthermia consider a setup like the one illustrated in Fig. 1a, where metallic nanoparticles flowing in a capillary generate local heat by excitation of LSP resonances when illuminated with a laser beam. The nanoparticles feature free rotational diffusion and therefore adopt all orientations with the same probability for the case of isotropic nanoparticles (see Fig. 1), highlighting the importance of considering the spectral thermal response of the nanoparticles for different orientations with respect to the light source. In the specific case of phototherapy against cancer, the temperature of the sample (either in vivo or in vitro) needs to reach temperatures between 42-48 degrees Celsius [45], namely, about 10 degrees above the average temperature of the biological medium, for the conditions and thermo-optical properties of the human tissue. However, it is necessary to exert a dynamic control of the temperature increases to avoid uncontrollably damaging the healthy surrounding tissue, as it would happen with continuous wave lasers. Pulsed lasers are widely used to avoid these issues, because tissues can suffer different processes such as photochemical damage, photothermal damage, photoablation or photo disruption as a function of the power density and the exposure time [46]. Thus, sudden and high thermal responses are desired at power densities as lower as possible. For photothermal therapies, this is achievable by considering more efficient nanoparticles in terms of light/heat conversion.

A thorough comparison of the photothermal spectral response of different particle shapes was performed. Electromagnetic and thermal simulations were carried out under plane wave Vis-NIR illumination unpolarized light, since as the light beam travels through tissues, multiple absorption and scattering phenomena take place, unpolarizing it partially.

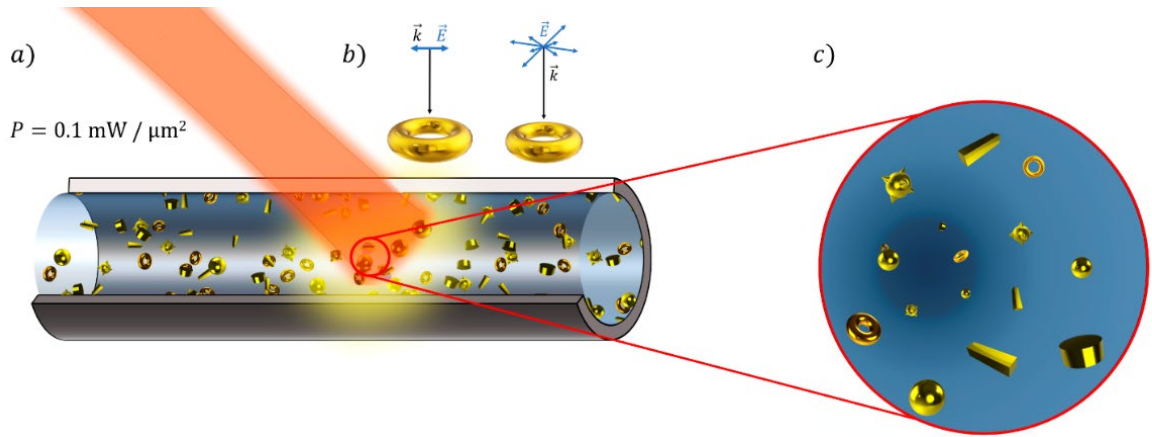

Figure 1: Standard thermofluidic application of this work. a) Studied nanoparticle shapes inside a fluidic channel. b) Illumination process of the system. c) Randomly oriented particle. Upon excitation by light, the effective temperature increase will correspond to the average temperature achieved by all possible orientations.

The geometries considered for this comparison are shown in Fig. 1 and include some of the most widely used shapes: spheres, spherical gold/silica core-shells, nanostars, disks, rectangular rods (similar results than rounded rods as shown in Fig. S1) and toroids. All particles have been designed considering a therapeutically allowed maximum dimension size of 120 nm and compared under the same conditions selecting the best design in terms of its potential heating capability. The particle designs are sought to provide the highest possible temperature for a given intensity. This would allow the nanostructures to generate heat efficiently even with minor incident power which is essential in photothermal therapy applications where human tissues induce an exponential decay of the light intensity [47]. Following previous studies [48,49], we consider the analytical solution of the classical heat equation for an initial analysis of the heating effects. A very good agreement between theory and simulation can be observed for spherical particles below 30 nm in radius. However, deviations are observed for larger particles in the regime where the particle size is not sufficiently small to fulfil the quasi-static approximation considered in equation (1). In this regime, the particle is influenced by a time dependent electric field which translates into a non-uniform electronic density motion, thus, reducing the magnitude of the resistive losses and consequently the electromagnetic absorption (see Fig. S2 and S3 for a systematic sizes study of both, electromagnetic absorption and temperature increment, respectively).

Figure 2 shows the photothermal spectral response of the complete set of nanoparticle shapes considering its steady (2a) and time dependent thermal response (2b) calculated in aqueous environment, irradiated at normal incidence for both, linearly polarized and unpolarized light with an intensity of 0.1 mW/µm$^2$, several orders of magnitude smaller than those considered in previous works [42,50,51] (see Fig. S4 for other excitation power densities). All particle shapes were optimized by changing the geometrical parameters of each shape (within a maximum dimension size of 120 nm) and selecting those that yield the greatest temperature increment. The optimal dimensions for each geometry offer a better thermal performance compared with other configurations of the same volume for the same structure (see Fig. S5 in the supplementary document). We then compare the photothermal behavior, focusing on the pros and cons that each shape offers A detailed physical discussion of the different responses can be found in the supplementary material.

In Fig. 2a, it can be seen that the disk, rod and toroid clearly outperform the sphere, core-shell and star shapes in terms of maximum temperature increase. Although the sphere is the most widely used geometry in biomedical applications (PTT, drug delivery, microfluidics, gene therapy and imaging, for instance) [52,53], partly due to the simplicity of fabrication, it shows a poor thermal response in comparison with other geometries. It displays a maximum temperature increment of $\Delta T \approx 6$ K, at a wavelength far from the biological windows, offering negligible tunability compared to the rest of the analyzed structures [54] and limiting its applicability to clusters, where collective effects play a definitive role. Meanwhile, the gold/silica core-shell although located closer to the more usable wavelengths of the first biological window (700-900 nm [15]), still shows a low temperature increase $\Delta T \approx 7$ K in contrast to the upper three shapes and shows subpar tunability [33], allowing its use only in the first and less beneficial window at the price of lower temperature increments (see figures S6 and S7 in the supplementary document). Finally, the nanostar geometry shows a similar behavior to the core-shell, with a mild temperature increment around 12 K at the same excitation wavelengths. However, its response is extremally sensitive to the number of cones and their sizes, hence, its fabrication is crucial and challenging to reach the desired photothermal response.

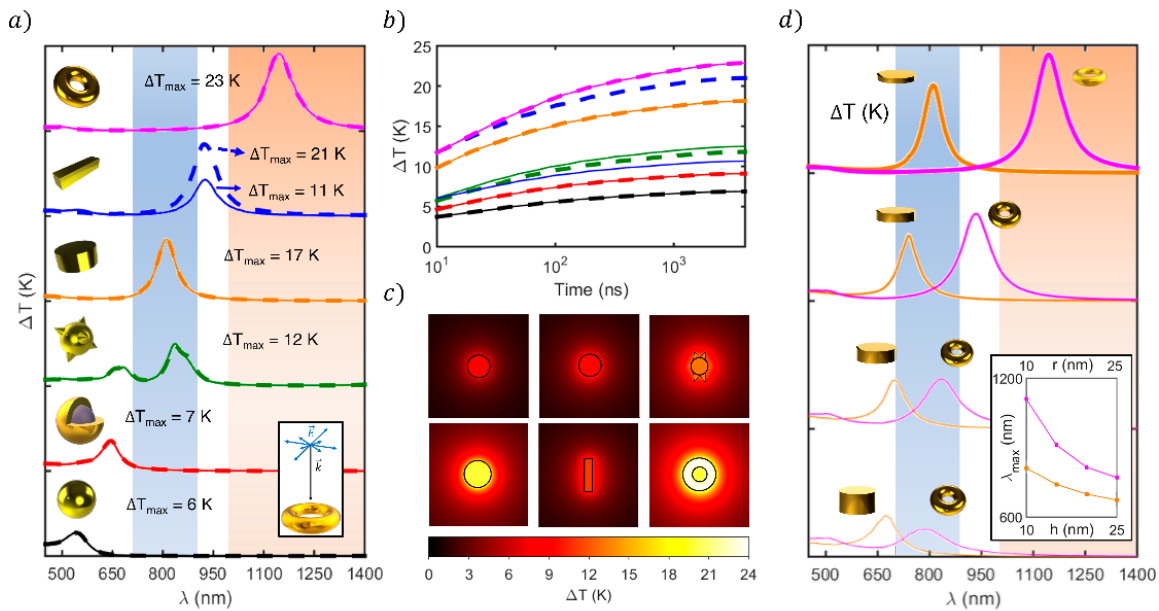

Figure 2: Thermal response of gold nanoparticle shape comparison in water for linearly polarized (dotted line) and unpolarized (continuous line) light at normal incidence. The sphere (in black) has a radius of 40 nm, core-shell (in red) have a core radius of 30 nm and a shell thickness of 10 nm. The core of stars (in green) has a radius of 30 nm with a cone height of 30 nm. The disk (in orange) radius is 50 nm, with a height of 10 nm. The dimensions of the rectangular rod (in blue) are 30x120x30 nm. Finally, the doughnut or toroid shape (in pink) has a main radius of 50 nm and a secondary radius of 10 nm. A) Optimal thermal response for each analyzed geometry considering a major dimension lower than 120 nm are plotted for temperatures between 0 and 25 K. The blue and reddish shadowed regions represent the first and second biological windows, respectively. B) Transient optimal point heating for the shapes shown in a). c) Temperature increase spatial distribution. D) Tunability comparison of disks and toroids between 0 and 25 K. The disk radius is fixed at 50 nm and the torus main radius at 50 nm. The disk heights are 10, 15, 20 and 25 nm and the torus secondary radii, 10, 15, 20 and 25 nm. Relative wavelength displacement is plotted in the inset.

On the other hand, the disk, rod and toroidal shapes show better thermal response in all aspects (magnitude, spectral tunability and response time). The disk shows a better performance in terms of heating however the optimal geometry requires an especially thin height of 10 nm with a radius of 50 nm to achieve the maximum temperature, resulting in a challenging fabrication process [55]. In the case of the nanorod, its thermal response displays a redshift with respect to the disk. Furthermore, a huge discrepancy between linearly polarized and unpolarized light responses exists. A linear polarization illumination (input electric field polarized along the greatest dimension of the rod) gives a temperature increment peak of 21 K, higher than the observed for the disk. However, for unpolarized light, a considerable reduction in the temperature increment can be observed, resulting in a lower maximum temperature increment of around 11 K, closer to the core-shell response.

This polarization dependence is a clear disadvantage in the performance of the rod against other particles, as the light beam usually becomes unpolarized when traveling through tissues, due to the multiple absorption and scattering phenomena. Finally, the last considered shape to improve thermal response in nanoparticles is a toroidal geometry, which offer the best performance, with an outstanding increment in maximum temperature increment of around 23 K, a 35% of increase (see Fig. S8 for temperature increments in other surrounding media) when compared with the second largest temperature increment in unpolarized illumination (17 K for the disk). As we will demonstrate later, the toroid also shows the greatest tolerance to rotations with respect to the incident beam, which guarantees a reliable performance in thermofluidic and photothermal therapy applications.

Figure 2b shows the results of the thermal time-dependent calculations at the optimal wavelengths depicted in Fig. 2a. It can be seen that the three worst particles require also longer times to reach the steady state (considered to be when 90% of the maximum temperature increment is reached, see figure S9 for further discussion), up to around 320 ns for the sphere and core-shell, the star having a larger time of 330 ns. In contrast, the three best heating candidates' shapes display shorter steady state reaching times more than 20% shorter (as short as 260 ns for disks and rods), furthering the case for these three nanoparticles. Furthermore, the toroidal particle also reaches its maximum temperature after a 300 ns illumination, being one of the three fastest heaters. This can be explained in terms of volume. As the optimal rods, disks and toroids have smaller volumes, they reach the steady-state in a shorter time. A clear way to summarize the thermal behavior of all studied shapes, is to consider its temperature spatial distribution (Fig. 2c). Disks and toroidal particles offer very high temperatures for unpolarized light and heat up the greatest surrounding region, while the rest of geometries show much lower heating capacity. Toroids also show high temperatures in the water region confined in the inner hole, while all geometries display their maximum temperature inside gold.

Another key feature when bringing these different heaters to practical applications is their spectral tunability, which has been investigated through variation of the geometrical parameters (see Fig. S10-18 for further information). This has been compared in Fig. 2d for disks and toroids. Starting from two designs that offer comparable thermal responses, the characteristic parameters (height for the disk and secondary radius for toroids) have been varied keeping the main radius constant for both geometries. It can be seen how for equivalent geometrical variations, we can get larger wavelength shifts. Moreover, in the case of the disk it is impossible to bring its thermal response to the second biological window (see inset of Fig. 2d). This clearly demonstrates the significantly better tunability of toroidal shaped nanoparticles. The comparison shown in Fig. 2, evidences the photothermal response superiority of the rod, disk and toroidal geometries, showing the last one features a better response in magnitude and spectral tunability (see Fig. S19 to visualize the corresponding study in air).

*3.2 Nanoparticle orientation dependance*

Another important aspect in photothermal applications, is the fact that the heating response of nanoparticles in a fluid strongly depends on the particle orientation with respect to the illumination source. Since nanoheaters will rotate freely around the fluid, the effective temperature obtained upon illumination will correspond to the average of the temperatures achieved by all possible orientations. Thus, particles with small thermal dependence to the incident beam - structure relative orientation will offer better effective performance. This aspect was considered in a previous theoretical work [42], under linearly polarized excitation, showing the absorption cross section for rods and toroids. In this work, we provide the angular average temperature value, appropriate to compare the heating performance, that the aforementioned nanoheater candidates can reach.

To clearly discern which of the three best heaters (rod, disk and toroid) stands out and why, we considered the three of them immersed in water and calculated how their relative orientation to the exciting light affects their thermal response. Figure 3 shows the temperature maps of unpolarized light excited nanoparticles for all possible 3D rotations with respect to the x and y axes, taking their optimal response orientation as the initial position (see Fig. S20 for a similar study in air). These rotations, seen as points on the unit sphere, are represented by $\theta$ and $\phi$ for the *x* and *y* axes respectively, as shown in the illustration of Fig. 3a. To extract the angular average temperature, we use a set of rotations equally distributed on the sphere surface, leading to a uniformly sparse set of points. This consideration is required to guarantee that all rotations are equiprobable and all points equally weighted.

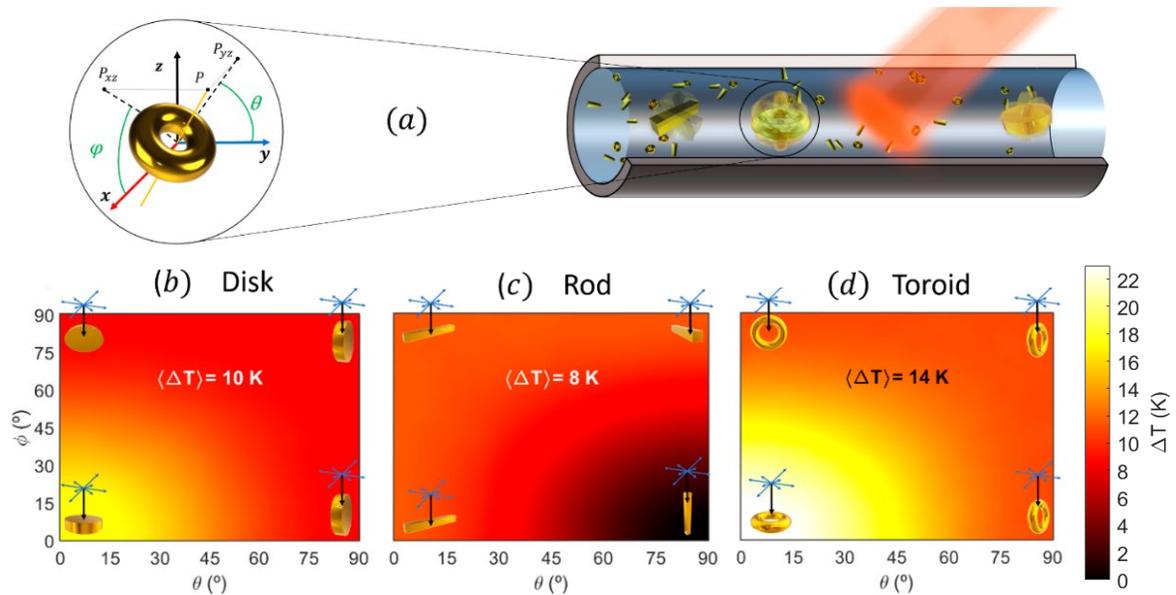

Figure 3: Thermal responses of the three best nanoheaters in Fig. 2 to rotations. a) Illumination and rotation outline. $\theta$ and $\phi$ are the rotations with respect to the *x* and *y* axes and the considered power density is 0.1 mW/µm². Maximum temperature increases for unpolarized light as a function of rotation angles: b) Disk; c) Rod; d) Toroid. The average temperature is shown for all cases.

Looking at the disk results in Fig. 3b, it can be seen how the temperature distribution shows a revolution symmetry in the colormap, as the revolution axis of the disk and the $z$ axis (beam direction) match in the initial configuration. Thus, considering the temporal evolution of the electric field vector, rotating the disk about the $y$ axis, and subsequently about the $x$ axis, has the same effect as rotating it about the $x$ axis and then about the $y$ axis. To facilitate the understanding of this figure, setting $\theta = 0º$ and then increasing the rotation in $\phi$ will mean that eventually the transverse length and the electric field will be eventually parallel, so that occasional resonances will be produced, resulting in a decrease in the mean value of the temperature increment of around 41% compared to the most favorable configuration ($\Delta T=17$ K). Thus, due to its symmetry, the disk exhibits a reasonably stable thermal response to rotation.

In Fig. 3c (rod), an obvious symmetry in the angular distribution of the temperature increment can be observed. At $\theta = 0°$, the major axis of the rod is orthogonal to the beam wavevector, and therefore the rotating electric field will occasionally align with this direction for any $\phi$ value, guaranteeing a constant thermal response. However, setting $\theta = 90°$ and $\phi = 0°$, the major axis is orthogonal to the electric field at any time, preventing favorable resonances and heating. In this case, as $\phi$ grows, the projection of the aforementioned major axis along the direction of the electric field vector increases, promoting enhancement of the thermal response. Figure 3c clearly shows that the rod displays a worse temperature angular distribution than the disk. Thus, the angular average temperature increment is significantly smaller (8 K), translating in a reduction of 62% with respect to the maximum achievable value ($\Delta T = 21\ K$). This is due to the fact that the electric field is oscillating in time, reducing the efficiency of the generated mode in the rod and consequently its thermal response.

The case of the toroidal particle, shown for the same angle configuration in Fig. 3d, exhibits weaker variations in the increase of temperature for any $\phi$ value upon rotation in $\theta$. A reduction of temperature increments is observed symmetrically for rotations in $\phi$ and $\theta$, achieving a stable value of around 12 K for a full 90º rotation in either $\phi$ or $\theta$. As in the case of the disk, as the electric field vector rotates over time, there always exists a configuration in which this vector is parallel to the toroid transversal length, allowing a continuous excitation of the corresponding mode. However, for each time in which the previous situation occurs, the opposite also takes place, with the electric field vector orthogonal to the toroid transversal length, avoiding the possible excitation mode. This leads to a compromise between both configurations, giving rise to a less time dependent excitation. The average temperature increment results in a reduction of 39% with respect to the most favorable configuration ($\Delta T = 23\ K$), proving that toroidal nanoparticles have a high tolerance to rotations and providing the highest angular average temperature increment compared to disks and rods.

### 3.3 Optimal Spectral Photothermal Response: A guide for toroidal particle design.

We have shown that toroidal particles offer the best nanoheating performance in our set of geometries, with minimal loss due to rotational diffusion. Here, we perform now, an exhaustive analysis of the dependence of the temperature increase with the geometrical features of this nanoparticle. This investigation provides a clear and direct guide to select the working spectrum and the specific dimensions of the torus depending on the specific application.

Figure 4 shows two color maps that correspond to the temperature increment produced by the toroid immersed in water (Fig. 4a) and the wavelength at which the optimal resonance occurs for each case (Fig. 4b). The corresponding study in air is shown in Fig. S21. As in the previous cases, possible phase changes in the gold nanoparticles have not been considered, since most applications involve an aqueous medium in which the maximum temperature increments are lower than 30 K for the considered power density. According to the aforementioned maximum size (120 nm), main radii lower than 50 nm have been examined. In addition, a secondary radius larger than 5 nm has been assumed. Below that value calculations become very speculative, because the deposition of metallic materials thinner than 10 nm is a huge manufacturing challenge, and the shape/size precision cannot be guaranteed [56,57].

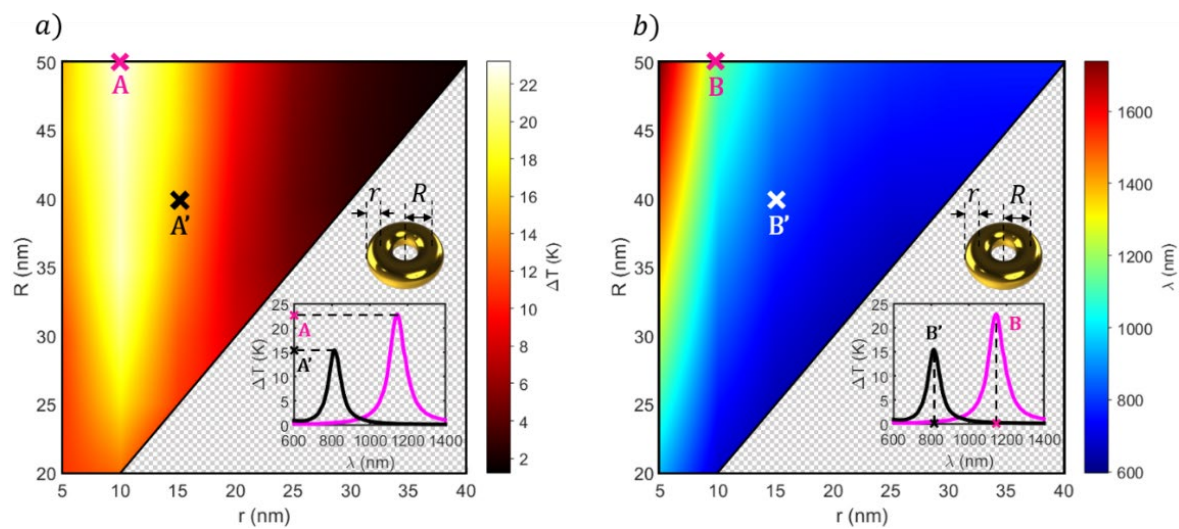

Figure 4: Thermal response of a fully tuned toroidal nanoparticle. a) Maximum temperature increases of toroid for main ($R$) and secondary ($r$) radius variations represented between 0 and 25 K. b) Corresponding excitation wavelength of the maximum thermal response as a function of the main and secondary radii. The points A and B correspond to the spectral maximum temperature increase (inset) for the toroid shown in Fig. 2, $R = 50$ nm and $r = 10$ nm. The points A' and B' correspond to other spectral maximum temperature increase for the $R = 40$ nm and $r = 15$ nm toroid.

The dependence of both parameters, temperature increase and excitation wavelength, with the dimensions of the nanodoughnut is directly related and follows the same pattern. Notice that the maximum increment of temperature is reached only when fixing the secondary radius around 10 nm. As we increase that parameter, the toroidal geometry gets closer to a sphere, and so does the increase in temperature. On the other hand, secondary radii smaller than 10 nm would lead to a reduced toroid cross section, hindering the support of strong electromagnetic modes that can produce high thermal responses. An opposite behavior appears when looking at the main radius, showing in general a growth of temperature increment for larger values. This can be explained in terms of the excited mode in the geometry. As the main radius increases, the toroid length grows shifting the excited mode towards longer wavelengths and enhancing the corresponding resistive losses.

## 4. Conclusions

We have performed a numerical investigation of the photothermal response of different gold nanostructures, demonstrating that the nanodoughnut shape displays an optimum performance with a temperature increment about 35% higher than the second-best heater shape, the nanodisk. The doughnut shape outclasses the temperature increments of the most widely used nanostructures, offering the largest heating areas, and also showing superior spectral tunability, supporting strong resonances within the therapeutic windows and particle sizes considering the recommended therapeutical limits for biomedical applications. We have also analyzed, the impact of nanoparticle rotational diffusion on its photothermal performance. The nano-doughnut shape features the highest average temperature increment and shows a high tolerance to rotation compared to other structures. For these reasons, it is foreseen as ideal candidate for photothermal applications. Finally, in view of the advantages offered by the toroidal shape, a simple and practical guide that can facilitate its design to act as an efficient heater in a desired spectral range has been provided.

We believe that the results and conclusions of this work highlight the promise of realistic and customized cancer treatments in a near future. This investigation could also help researchers to optimize the thermal performance of other fluidic systems, such as those using phase changing material heating, wettability or oil heating. Although in this study the maximum dimension has been limited to 120 nm following the most common recommendations in biomedical applications, Fig. 4 suggests that larger nanodoughnuts can also be exceptional candidates in other spectral regions out of the biological windows, making them available for other applications. Thus, our investigation, although focused on biological applications, could be extended to other areas that may need a careful selection of the nanoheater material and its consequent spectral photothermal optimization.

## 5. Methods

Metal nanoparticles can efficiently generate heat under optical excitation, especially in the plasmon resonance regime. The heat generation process involves not only absorption of incident photons, but also the heat transfer from the nanoparticle to the surrounding medium. A very simple mechanism takes place: the incident electric field strongly drives mobile carriers of the metal, heating the material owing to the energy gained by these carriers. Then, heat diffuses away from the nanostructure and leads to a temperature increase of the surrounding medium. In the absence of phase transformations, the temperature distribution around optically stimulated nanoparticles are described by the usual heat transfer equation:

$$\rho(\boldsymbol{r})c(\boldsymbol{r})\frac{\partial T(\boldsymbol{r},t)}{\partial t} = \nabla k(\boldsymbol{r})\nabla T(\boldsymbol{r},t) + Q(\boldsymbol{r},t) \qquad (4)$$

where r and t are the position and time, T(r,t) is the local temperature and the material parameters $\rho(r)$, $c(r)$ and $k(r)$ are the mass density, specific heat and thermal conductivity, respectively. The function Q(r,t) represents the energy (heat) source coming from light dissipation (electromagnetic losses). The solution of this equation has a transient state, and after a characteristic time, it reaches its steady state. Thermal processes in metals are fast, which means that a steady state is rapidly reached for typical incident powers and metal nanoparticle dimensions, similar to those used in nanomedicine. To obtain the electromagnetic losses, the system of Maxwell's equations including appropriate boundary conditions must be written, which in the case of complex topologies, such as core-shell or stars geometries, must be solved numerically. In our case, the whole process of light absorption and subsequent heat transfer between the nanostructure and the surrounding medium has been modelled by means of finite element simulations. For easy implementation and reliability of the solution, we have chosen Comsol Multiphysics 5.6, which provides state-of-the-art routines to solve partial differential equations. In our simulations, we have assumed that particles are illuminated with circularly polarized light to emulate the more realistic unpolarized illumination. Unpolarized light can be simulated by just solving the problem for two orthogonal input electric fields, but this doubles the simulation time. Thus, it can be seen that circular polarization gives equivalent results while being computationally more efficient (see Fig. S22 for an extended discussion). The electromagnetic losses from the electromagnetic waves in the nanoparticles as the only heat source. Furthermore, we have assumed that the electromagnetic cycle time is short compared with the thermal time scale (adiabatic assumption). To consider heat dissipation in our simulation region, we used a heat flux node across the outer boundaries, considering a heat transfer coefficient, dependent on the geometry and the ambient flow conditions. The heat transfer coefficient $h$ can often be estimated by dividing the thermal conductivity of the fluid by a length scale.

**Funding.** Ministerio de Economía, Industria y Competitividad, Gobierno de España (PGC2018-096649-B-I).

**Acknowledgments.** Authors would like to thank A. Franco and C. R. Crick for the interesting discussions. We gratefully acknowledge financial support from Spanish national project (No. PGC2018-096649-B-I), the UK Leverhulme Turst (Grant No. RPG-2018-384), UK-EPSRC (EP/J003859/1) and Imperial College Europeans Partner Fund grant. J. G-C. thanks the Ministry of science of Spain for his FPI grant. G. S. thanks the Ministry of education for his collaboration grant and P.A. acknowledges funding for a Ramon y Cajal Fellowship (Grant No. RYC-2016-20831).

**Disclosures.** The authors declare no conflict of interest.

**Supplemental document.** See Supplement 1 for supporting content.

# Supplementary material

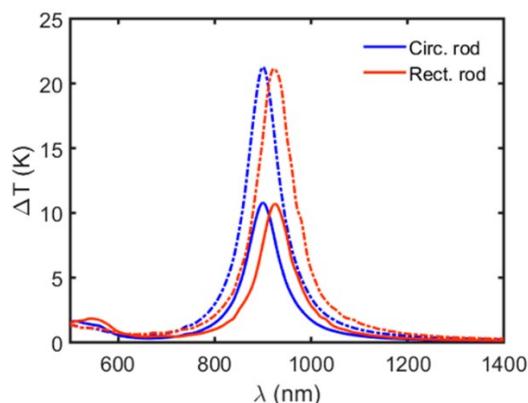

*Figure S1: Spectral response comparison of circular (blue) and rectangular (red) rod in water for linear (dotted line) and circular (continuous line) polarizations. The dimensions of the rectangular rod are 120 nm length with a section length of 30 nm. The circular rod has the same length and the needed radius to remain the volume constant.*

Figure S1 has been constructed to justify the election of the rectangular rod. The circular and rectangular rod thermal responses show similar maximum temperature increases for both incident polarizations, reducing its thermal efficiency for circular polarization as shown above. It can be seen that rectangular rod response is slightly redshifted with respect to the circular one making them equivalent. As the circular geometry corresponds to the maximal cross section for a fixed volume, the aspect ratio (minor length over major length) is increased with respect to the squared rod, explaining the observed blueshift.

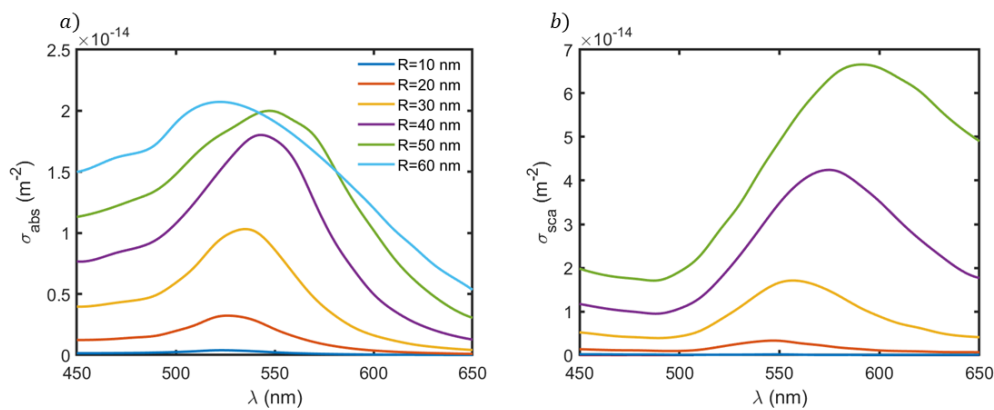

*Figure S2. Absorption (a) and scattering (b) cross-sections of an isolated gold sphere in water for a set of radii.*

In figure S2, the scattering and absorption cross-sections for a gold sphere in water are shown. It can be seen, the well-known redshift in the scattering spectra as a result of the diameter increase. The absorption cross section also presents a slight redshift for the smallest particles.

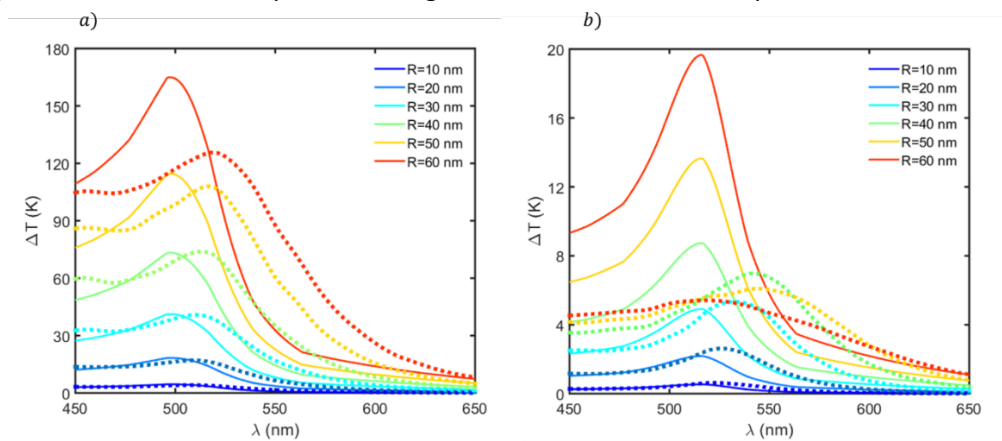

*Figure S3: Simulated spectral thermal response of a sphere (dotted line) compared with the obtained according to the expressions used by Govorov and co-workers (solid line) in air (a) and in water (b). Radii between 10 and 60 nm have been studied.*

Figure S3 shows the spectral thermal response of a sphere in air and in water for radii between 10 and 60 nm. In the last case, the obtained temperatures are much lower than in air due to the higher thermal conductivity of water. Attending to the Govorov calculations, the increase of temperature grows with the radius, keeping the resonance wavelength almost constant for both surrounding fluids and being redshifted in water with respect to air as can be expected from Mie theory. Although our simulations in air show a good agreement between curves for lower radii, it can be seen how a huge discrepancy appears for higher values, being the simulated maximum of temperature lower than the corresponding one given by the Govorov's expressions.

This significative difference can be explained in terms of the relative size of the excitation wavelength and the particle. For higher radii, the quasi-static approximation gets weaker, avoiding the description of the sphere as a perfect electric dipole. Furthermore, a possible thermal explanation could come from the resistive losses which are concentrated in a smallest and superficial region reducing the temperature increase at the particle in such a way that the relation between maximum temperature increment and particle size becomes more complex. Spheres show a poor spectral tunability for both surrounding mediums. In the following, we will show a comparison of all the analysed geometries (gold/silica core-shells, stars, disks, rods and toroids) in air together with some of the most representative nanoparticle designs for circular polarization, demonstrating the impact that every topological parameter has on its thermal behaviour.

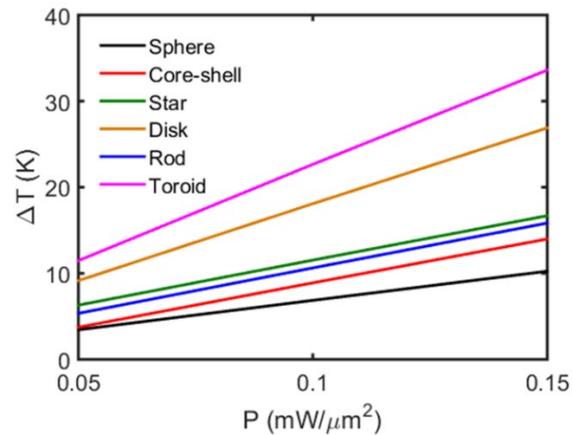

*Figure S4: Maximum temperature increase of the selected structures in Fig. 2 illuminated with circular polarized light as a function of the incident power density.*

Figure S4 shows how the power density affects the thermal efficiency of the nanostructures, showing that the temperature can be enhanced by increasing the power density. Noticeable differences between curves appears. Focusing on the slopes, it can be observed that disks and toroids present the fastest growth of temperature for higher power densities being the toroid one superior in contrast with the rest of structures. This can be explained in terms of the particles volume. Since spheres, core-shells, stars and rods are bigger than toroids and disks for the optimal dimensions shown in Fig. 2, they present a weaker growth with respect to the power density.

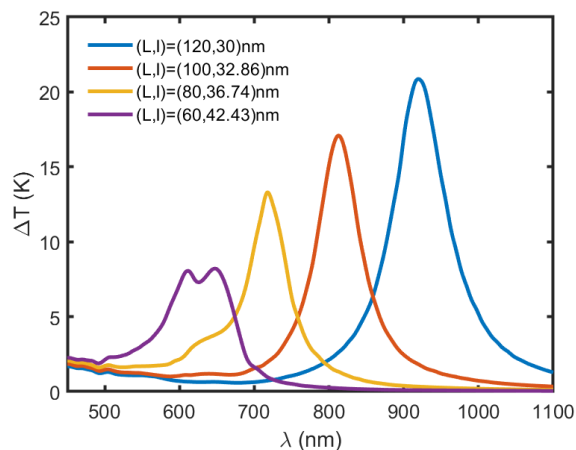

Figure S5. Spectral thermal response for a set of gold rods with identical volumes. The largest dimension is limited to 120 nm.

FigureS5 shows the spectral thermal response of gold rods with different sizes remaining their volume constant. It can be seen how the maximum temperature increment is decreased as the largest rod dimension reduces. This can be explained in terms of volume: since it must be constant, the shortest dimension of the rod increases for a decreasing larger dimension reducing the resistive losses and the rod heating ability consequently.

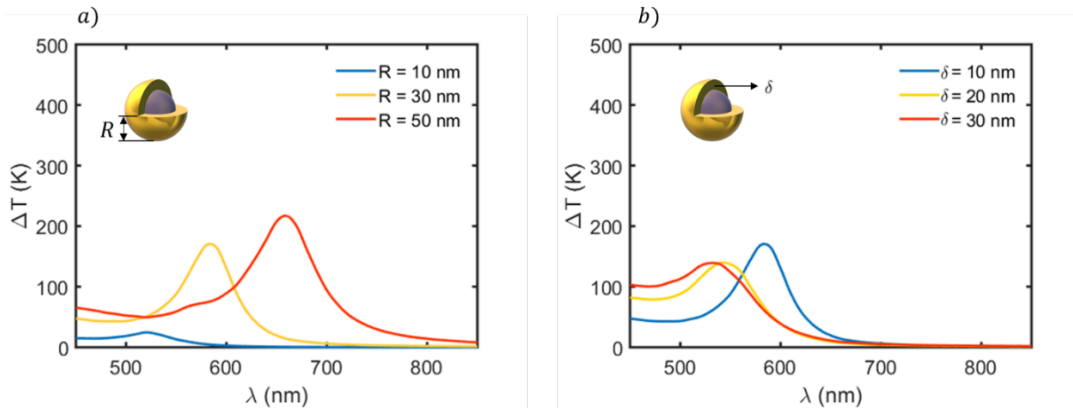

*Figure S6: Gold/silica core-shell simulation for variations in the core radius (R) and the shell thickness (δ) in air. a) Core-shell thermal response for core radius between 10 and 50 nm for a given shell thickness of 10 nm. b) Core-shell thermal response for shell thickness between 10 and 30 nm for a fixed core radius of 30 nm.*

Figure S6 shows the spectral thermal response of a core-shell structure and how the silica core makes it redshift with respect to the single material sphere (Fig. S3), pushing the system closer to the biological window. In Fig. S6a, the effect of the core radius is clearly demonstrated: as the core size increases for a fixed shell thickness, the spectral response is redshifted and the maximum temperature grows since the shell expansion carries to an increment of the resistive losses. On the other hand, if the core radius is fixed, the structure seems more like a sphere for thicker shells, giving rise to a reduction in the maximum temperature and a blueshift of the spectral response.

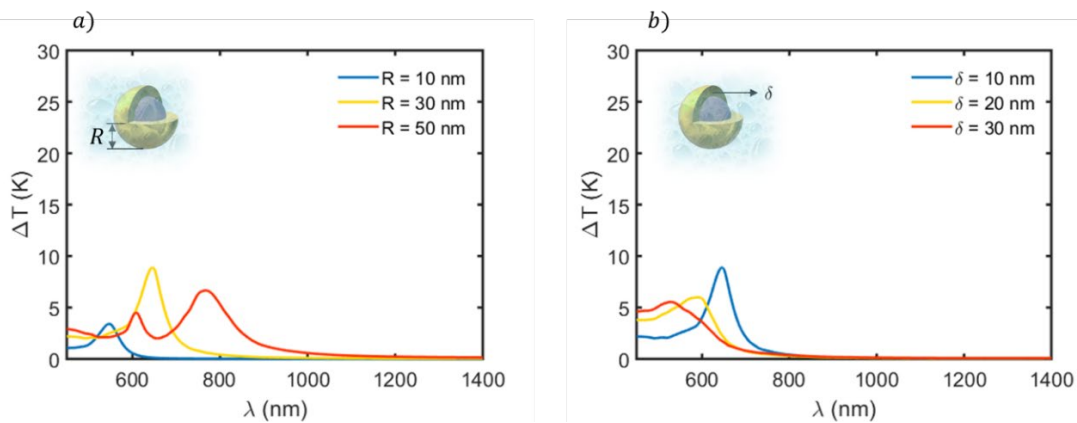

*Figure S7: Gold/silica core-shell simulation for variations in the core radius (R) and the shell thickness (δ) in water. a) Core-shell thermal response for core radius between 10 and 50 nm for a given shell thickness of 10 nm. b) Core-shell thermal response for shell thickness between 10 and 30 nm for a fixed core radius of 30 nm.*

Figure S7 shows the spectral response of a core-shell in water for different core radii and shell thicknesses. As in the case of air, in Fig. S7a, it is shown how as the core size increases for a given shell thickness, the spectral response is redshifted for increasing core radius. Conversely, as plotted in Fig. S7b, for a fixed radius, the structure seems more like a sphere for thicker shells, leading to a reduction in the maximum temperature and a blueshift of the spectral response.

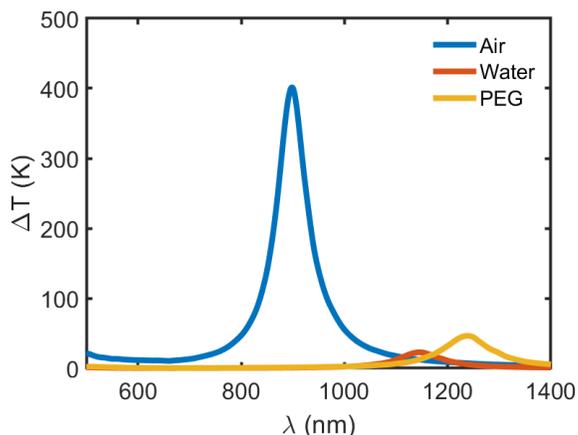

*Figure S8: Comparison of the spectral thermal response of a toroid immersed in air (blue), water(red) and PEG (yellow) for circular polarization. The dimensions of the toroid are $(R, r) = (50, 10) nm$.*

Figure S8 shows the thermal spectral response of a toroid immersed in three different fluids: air, water and PEG. It can be seen the huge discrepancies between the optimum temperature increase for the three media, reaching temperatures of 400 K in air, 23K in water and 45K in PEG approximately. This can be explained in terms of their thermal conductivity: 0.02 W/mK for air, 0.6 W/mK in water and 0.3 W/mK for PEG. Thus, larger temperature increments are obtained for less thermal conductive media. Another remarkable variable to consider is the refractive index, the PEG being the most refringent. These contrasts lead to the redshifts in Fig. S8.

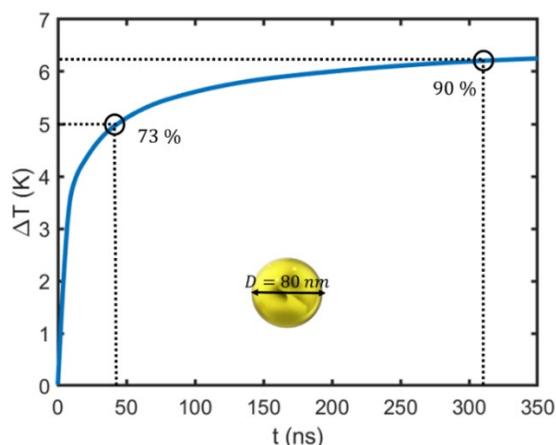

*Figure S9: Transient thermal state for an isolated gold sphere of 80 nm diameter in water illuminated with unpolarized light at 540 nm.*

Figure S9 shows the thermal transient state of a gold sphere in water illuminated at 540 nm. As can be seen, if we consider the criterion followed in this work, the sphere takes 310 ns to reach the 90% of the stationary temperature. However, this value variates depending on the selected criteria. For instance, if we consider a logistic growing, then the characteristic time will be reached at 73% of the stationary temperature approximately. Thus, the characteristic time would be 45 ns as seen in the figure, which fits with the estimated values from $t \cong D^2/a$ considering the diameter as characteristic length. Thus, the temporal values depend on the selected criterion without affecting the calculations suitability. In our work, the 90% of the stationary temperature was selected as the conductive criterion.

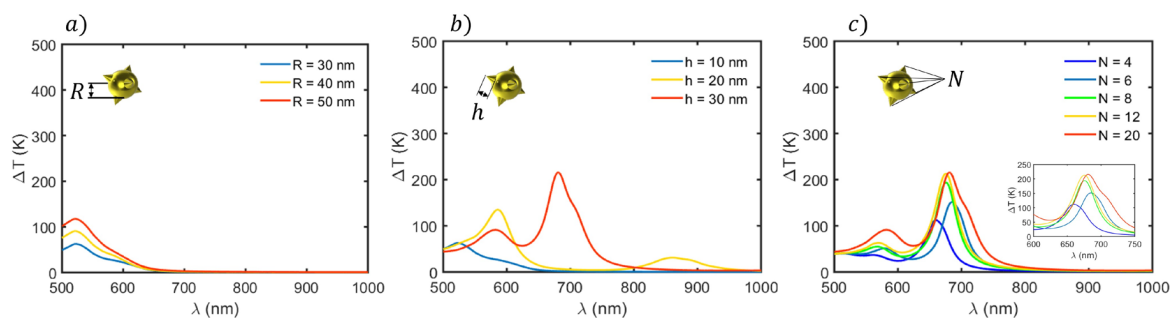

*Figure S10: Star simulations for a variating core radius (R), cone height (h) and number of cones (N) in air. a) Thermal spectrum for a star geometry with 20 cones of height 10 nm. b) Thermal spectrum for a star with 20 cones and a core radius of 30 nm. c)Thermal spectrum of a star geometry with a core radius of 30 nm and a variable number of cones of 30 nm height.*

Figure S10 demonstrates the tunability of nanostars, an alternative approach to improve the thermal response of nanoparticles. Attending to Fig. S10a, we can see how the thermal response is weakly redshifted with respect to the core-shell structure and a secondary resonance appears below 600 nm. The dual resonance observed for the nanostar is due to this intricated geometry: the smaller peak corresponds to a resonance of the spherical core, while the most prominent corresponds to a whole-structure resonance. As shown in Fig. S10b, if the cones height is reduced, the second resonance weakens and the nanostar approaches the single sphere thermal response.

The highest temperature increment obtained with this structure is 200 K. Although the nanostar structure gives a better thermal response than in the case of core-shell and sphered particles for an equivalent temperature increment and wavelength region, the structure is highly regular and the response is extremally sensitive to the number of cones and its sizes as can be seen in Fig. S10c. Hence, the nanostar fabrication requires high precision, becoming quite challenging. Due to this, the resulting structures are often irregular in such a way that calculations considering regular geometries are speculative.

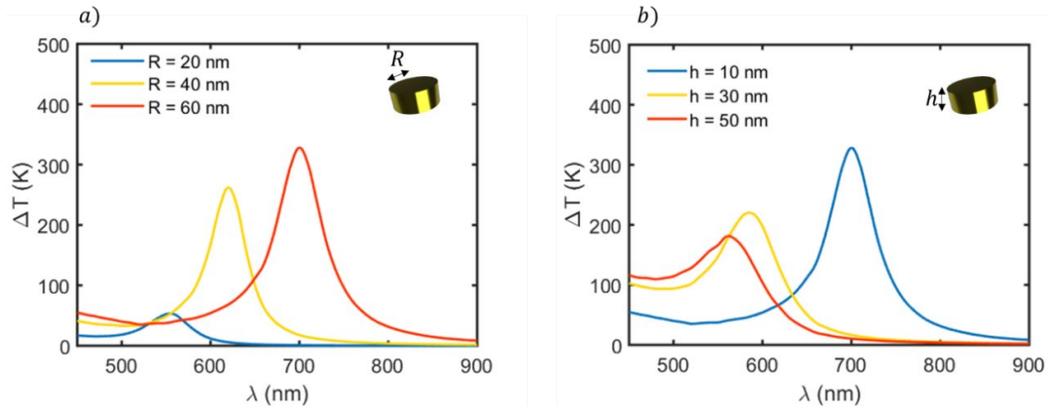

*Figure S11: Disk simulations for a variating radius (R), and height (h) in air. a) Thermal spectrum for a disk geometry of height 10 nm. b) Thermal spectrum for a disk with 60 nm radius.*

Figure S11 shows the spectral thermal response of the nanodisk structure for a set of representative dimensions. It can be seen in Fig. S11a how, as the radius increases, the thermal response is redshifted and higher temperature increments are obtained. The reason for this is that a radius enlargement leads to a growth in the electronic motion amplitude, giving rise to such a redshift and temperature increment. In addition, as Fig. S11 plots, a reduction of the disk height produces an augmentation of the temperature increment, as the structure resistance is increased. In case of the optimal disk (10 nm height, 60 nm radius), the maximum temperature enhancement is 320 K approximately.

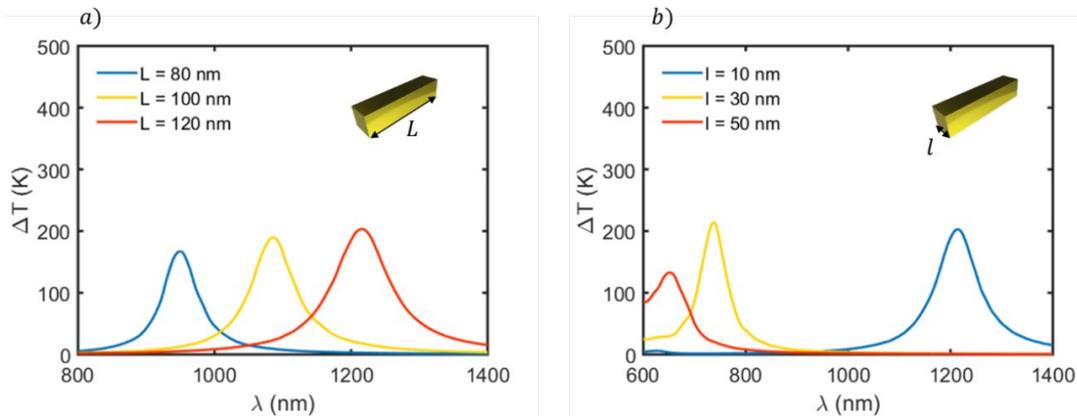

*Figure S12: Rod simulations for a variating major length (L) and section length (l) in air. a) Thermal spectrum for a rod geometry of l=10 nm. b) Thermal spectrum for a rod of L=120 nm.*

Figure S12 shows the spectral thermal tunability of the rectangular rod geometry. In Fig. S12a, it can be seen how as the major dimension increases, electrons within the material will oscillate with higher amplitudes, resulting in a larger excitation wavelength. In turn, as the section size increases and approaches the major dimension, the geometry will be akin to a sphere, resulting in a resonance blueshift (Fig. S12b). The rod response to circularly polarized light can be explained by considering the rotating nature of the circular polarization: as the electric field rotates, the major dimension of the rod and the electric field will only concur at specific times, at which a dipolar mode for the adequate wavelength is generated. Conversely, at successive stages, they will be crossed in such a way that the previous mode will disappear, resulting in a lower maximum temperature increment of around 200 K.

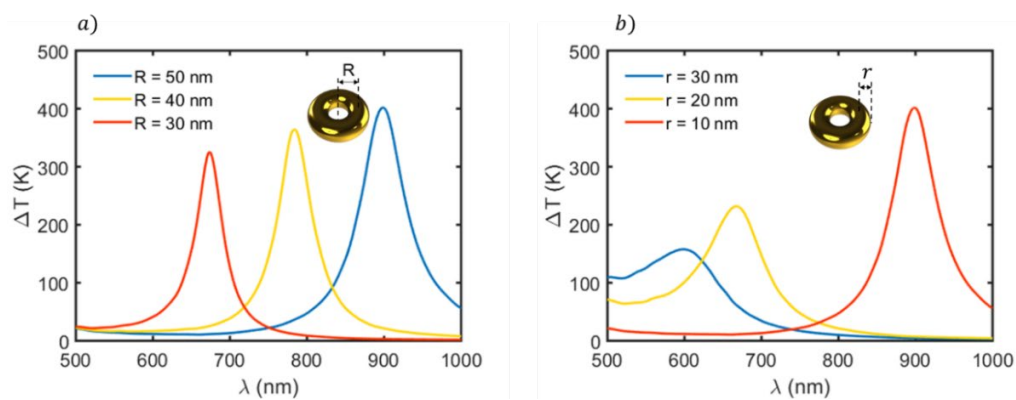

*Figure S13: Toroid simulations for a variating main and secondary radius (R and r) in air. a) Thermal spectrum for a doughnut of 10 nm secondary radius. b) Thermal spectrum for a doughnut of 50 nm main radius.*

Figure S13 shows the spectral tunability of the toroid which is the last studied shape to improve the nanoparticle thermal response. A maximum temperature increment of around 400 K is obtained for the optimal toroidal geometry (50 nm major radius, 10 nm minor radius). In Fig. S13a it can be seen, how as the major radius increases, temperature increments occur for higher excitation wavelengths, as a result of the interaction between the two toroid sides. If the main radius grows, the induced dipoles move away leading to a strong redshift which gives the toroid a huge tuning capability.

On the other hand, as plotted in Fig. S13b, if the secondary radius increases, the opposite effect can be observed: as this radius increases, the induced dipoles get closer, resulting in a significant blueshift. This clearly demonstrates the superior tunability of toroidal shaped nanoparticles over the rest of structures. The previous study has been performed considering the nanostructures immersed in water to expose the effect that the topological parameters have on the spectral thermal response of each geometry.

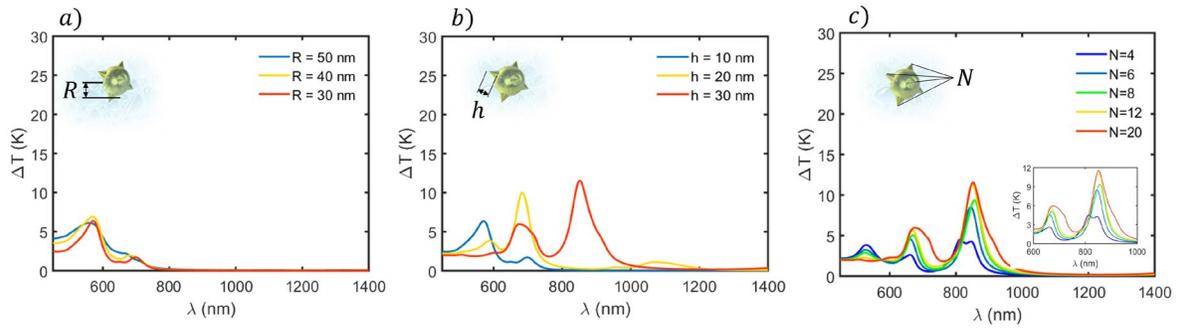

*Figure S14: Star simulations for a variating core radius (R), cone height (h) and number of cones (N) in water. a) Thermal spectrum for a star geometry with 20 cones of height 10 nm. b) Thermal spectrum for a star with 20 cones and a core radius of 30 nm. c)Thermal spectrum of a star geometry with a core radius of 30 nm and a variable number of cones of 30 nm height.*

Figure S14 shows the tunability of nanostars in water. Attending to Fig. S14a, we can see how the thermal response, generally, is weakly redshifted with respect to the core-shell structure. The evolution of the thermal response upon modification of the topological parameters exhibits a similar behaviour to the air medium case.

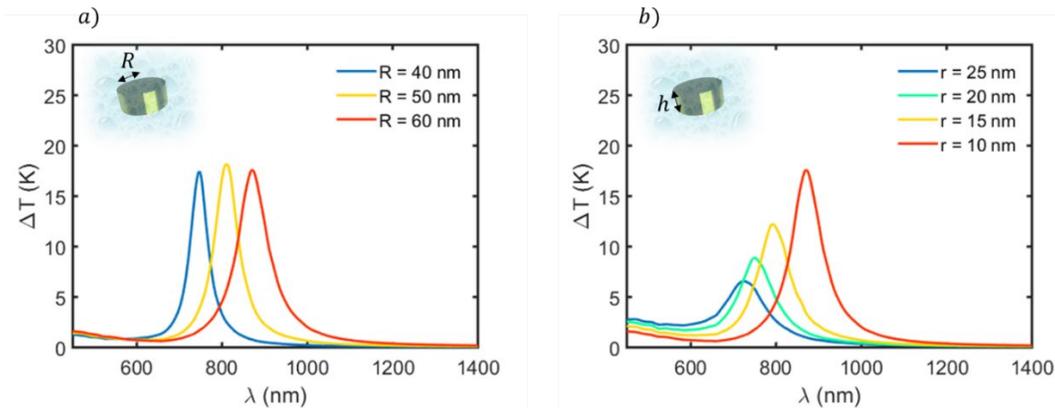

*Figure S15: Disk simulations for a variating radius (R), and height (h) in water. a) Thermal spectrum for a disk geometry of height 10 nm. b) Thermal spectrum for a disk with 60 nm radius.*

Figure S15 shows the tunability of disks in water. Attending to Fig. S15a, we can see how the thermal response shows similar temperature increments while the usual redshift is observed for larger radii. Similarly to the air case, reduction of the disk height result in a better thermal performance as can be seen in Fig. S15b.

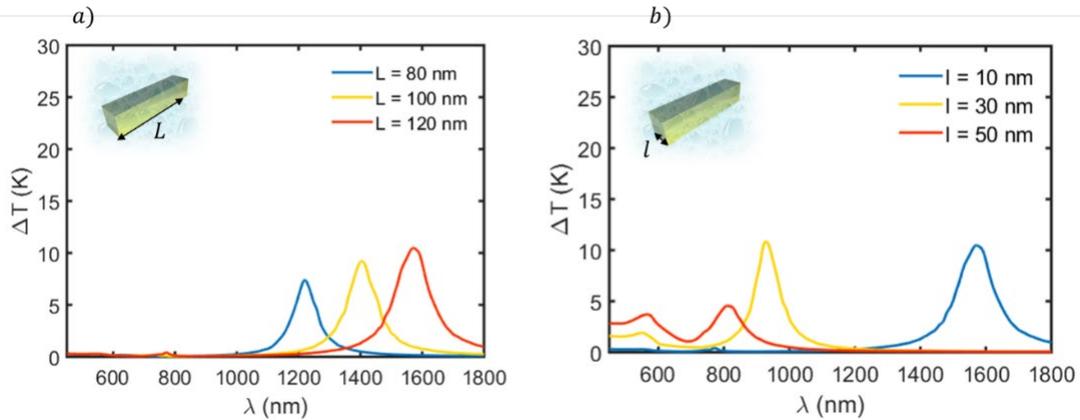

*Figure S16: Rod simulations for a variating major length (L) and section length (l) in water. a) Thermal spectrum for a rod geometry of l=10 nm. b) Thermal spectrum for a rod of L=120 nm.*

Figure S16 shows the spectral thermal tunability of the rectangular rod geometry in water. In Fig. S16a, although redshifted, the thermal spectrum shows a similar behaviour to the air calculations when the structure dimensions are modified. However, as depicted in Fig. S16b, a new optimal configuration, with a 30 nm section length, appears.

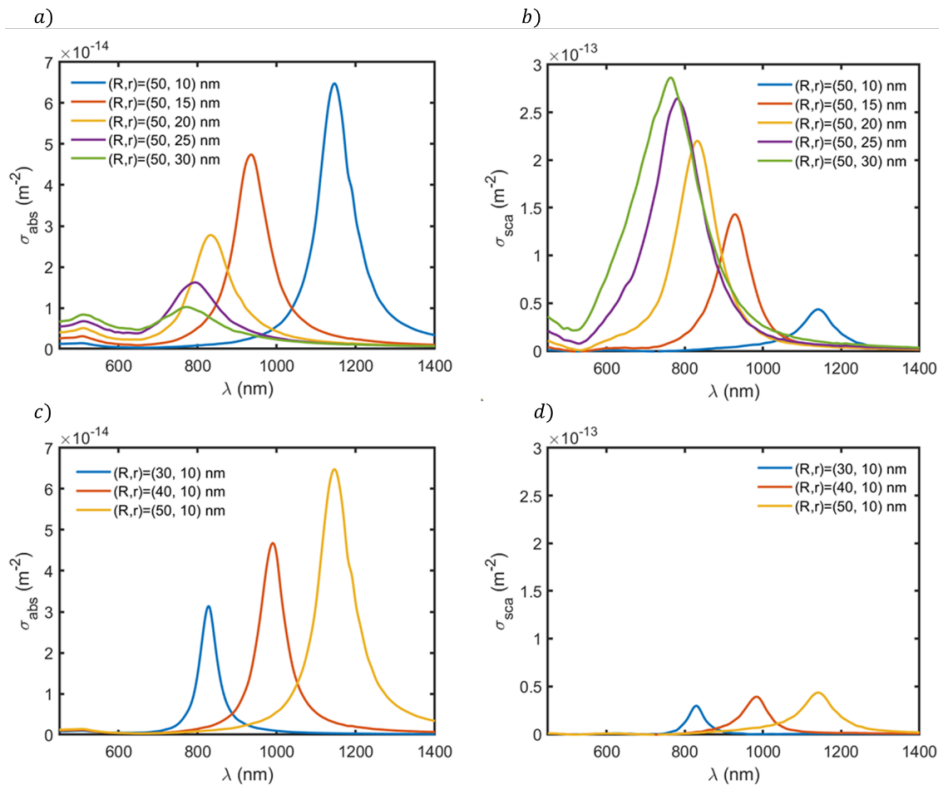

*Figure S17. Absorption (a, c) and scattering (b, d) cross-sections of gold nanodoughnuts in water for different sizes, ranging from 30 to 50 for the main radii and 10 to 30 nm for the secondary one.*

Figure S17 shows the absorption and scattering cross-sections for a set of gold nanodoughnuts. In (a) and (b), the main radius remains constant for a variating secondary radius revealing an interesting behavior: for increasing secondary radius, the spectral response is redshifted and becoming the toroid a better absorber, which fits with the temperature results shown in Fig. S18. In contrast, the scattering cross-section is reduced for growing secondary radius. On the other hand, in (c) and (d), the same pattern can be seen in the absorption cross-section: the absorption increases as the main radius grows. Besides, the scattering cross-section remains almost constant, disregarding the well-known redshift.

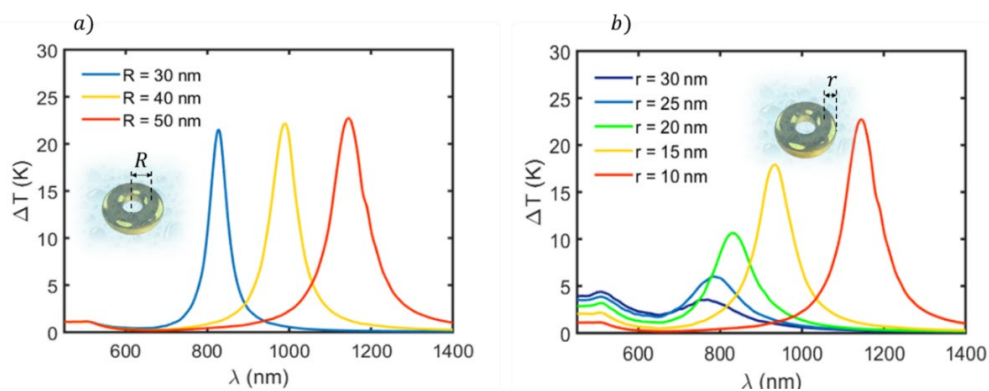

*Figure S18: Toroid simulations for a variating main and secondary radius (R and r) in water. a) Thermal spectrum for a doughnut of 10 nm secondary radius. b) Thermal spectrum for a doughnut of 50 nm main radius.*

Figure S18 shows the tunability of toroids in water exhibiting a behaviour fully parallel to its air counterpart. Focusing on the target wavelength region (second biological window, 1000-1400 nm), it can be seen how the 50 nm major radius and 10 nm minor radius configuration is the only one capable of reaching that spectrum.

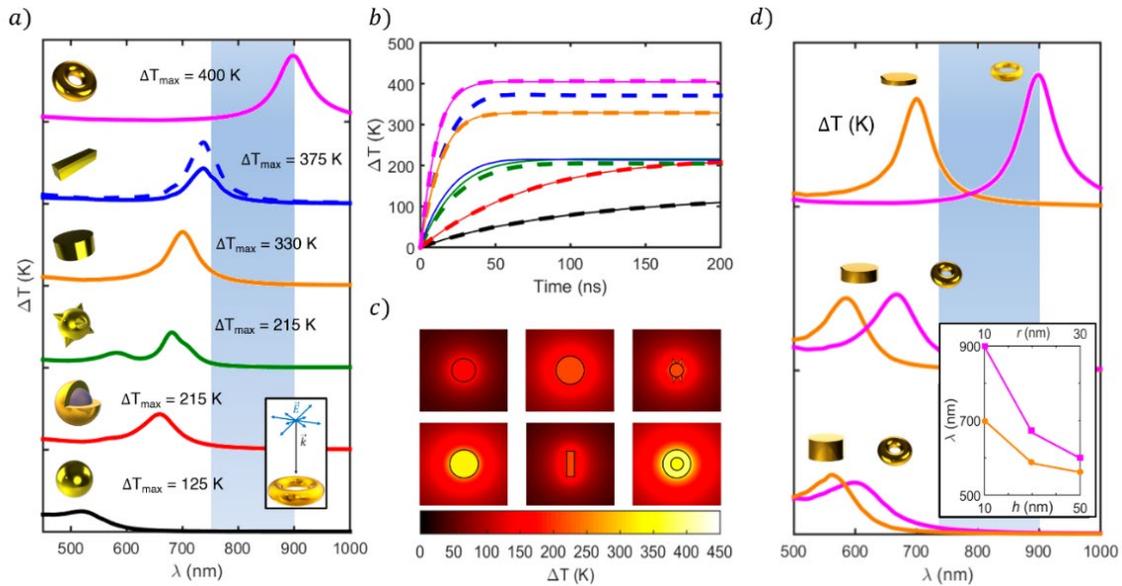

*Figure S19: Gold nanoparticle shape comparison in air for linear (dotted line) and circular (continuous line) polarizations at normal incidence. The thermal responses of spheres (in black), gold/silica core-shells (in red), stars (in green), disks (in orange), rods (in blue) and toroids (in pink) are represented. a) Optimal thermal response for each analysed geometry considering a major dimension lower than 120 nm are plotted for temperatures between 0 and 500 K. The sphere has a radius of 60 nm, cores-shells have a core radius of 50 nm and a shell thickness of 10 nm. The core of stars has a radius of 30 nm with a cone height of 30 nm. Disks radius is 60 nm, with a height of 10 nm. The dimensions of the rectangular rod are 30x120x30 nm. Finally, the toroid has a main radius of 50 nm and a secondary radius of 10 nm. The blue shadowed region represents the first biological windows. b) Transient optimal-point heating for the shapes shown in a) in the corresponding colours. c) Temperature increase spatial distribution for the studied shapes in a). d) Tunability comparison of disks and toroids represented between 0 and 500 K. The disk radius is fixed at 60 nm and the torus main radius at 50 nm. The disk heights are 10, 30 and 50 nm and the torus secondary radii, 10, 20 and 30 nm. Relative wavelength displacement is plotted in the inset.*

Figure S19 shows the photothermal spectral response of the set of the different optimized structures considering its steady (a) and time dependent thermal response (b). Particles are in air being irradiated at normal incidence for circular and linear polarizations. As in Fig. 2, the three upper shapes (disk, rod and toroid) clearly outclass the lower three (sphere, core-shell and star) in terms of temperature generation. Spherical geometries are the most extended particles due to its simplicity, however, they show the worse thermal response offering temperature increments of $\Delta T \approx 125$ K as can be seen in Fig. S19a. These increments reach their maximum in a wavelength far from the biological windows. Meanwhile, the gold/silica core-shell although located in a more usable wavelengths; still shows a poor temperature increase $\Delta T \approx 215$ K compared with the upper three shapes. Finally, the nanostar geometry shows a similar behaviour to the core-shell, with an optimal temperature increment of around 215 K at the same excitation wavelengths.

On the other hand, as in the case of water, the disk, rod and toroidal shapes demonstrate higher temperature increases with respect to the others. The rod response exhibits a redshift compared with the disk. As can be seen, a huge discrepancy between linear and circular polarization responses appears. The linear polarization illumination (input electric field polarized along the greatest dimension of the rod), gives a temperature increment peak of 375 K, higher than the disk one. However, for circular polarization, a considerable reduction in the temperature increment can be observed, resulting in a maximum temperature increment of around 200 K, closer to the core-shell response. Finally, in the case of the toroidal geometry, a maximum temperature increment of around 400 K can be seen for the optimal geometry, equivalent for both linear and circular polarizations.

Figure S19b shows the results of the thermal time-dependent calculations at the optimal wavelengths depicted in Fig. S19a. The three worst particles (sphere, core-shell and star) require also longer times to reach the steady state, up to around 200 ns for the sphere and core-shell, the star having a significantly shorter time of 100 ns. In contrast, the three best heating candidates' shapes display steady state-reaching times as short as 50 ns, furthering the case for these three nanoparticles. Furthermore, the toroidal particle displays a rapid heating, reaching its maximum temperature after a short 30 ns illumination, being the fastest heater.

The temperature spatial distribution shown in Fig. S19c briefly summarizes the thermal behaviour of the studied shapes. As it is shown, disk and toroidal particles offer very high temperatures for circular polarization and heats up the greatest surrounding volume, compared with the rest of structures. Toroids also show great temperatures in their inner holes, outside the material while all geometries display their maximum temperature inside gold.

Another essential aspect to be considered is the particles spectral tunability. This has been compared in Fig. S19d for disks and toroids, which are the two best candidates due to their excellent thermal response and polarization non-dependence. Starting from two designs with similar thermal responses, the characteristic parameters (height for the disk and inner radius for toroids) have been varied. It can be seen how for equivalent geometrical variations, we can get larger wavelength shifts demonstrating the exceptional tunability of toroids.

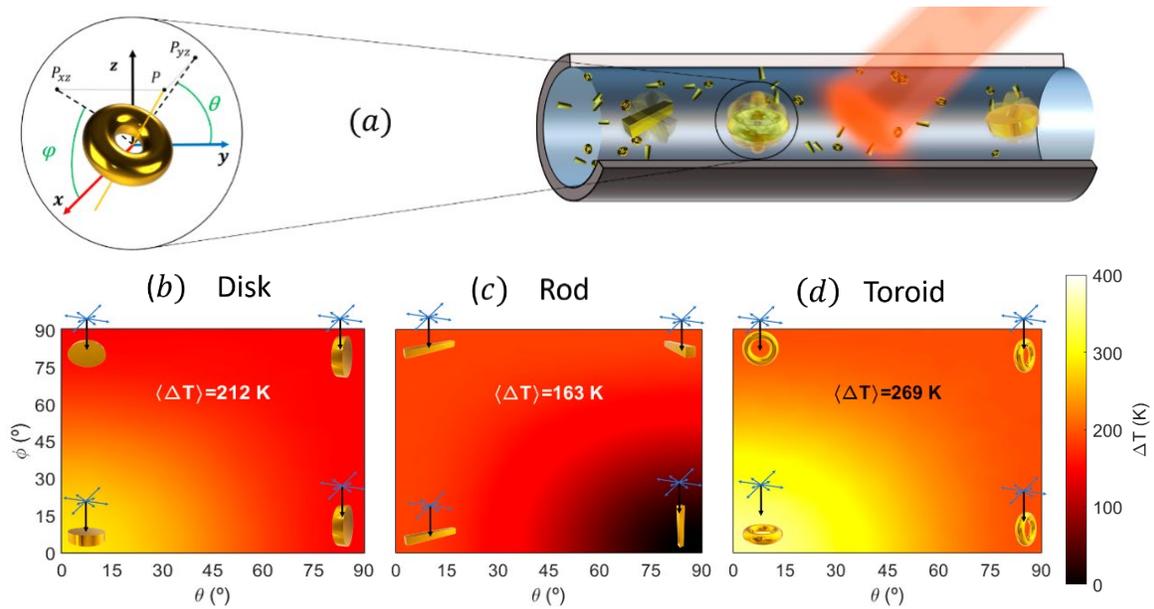

*Figure S20: Thermal responses of the three best nanoheaters in Fig. S19 to rotations. a) Illumination and rotation outline. $\theta$ and $\phi$ are the rotations with respect to the x and y axes and the considered power density is $0.1\ mW/\mu m^2$. Maximum temperature increase for linearly polarized and unpolarized light as a function of rotation angles: b) Disk; c) Rod; d) Toriod. The average temperature is shown for all cases. Top insets: linear polarization, bottom insets: unpolarized light.*

In Fig. S20, the implications of the particle orientation in the thermal response of the aforementioned structures are clearly demonstrated in air. The disk colormap in Fig. S20b, shows a certain revolution symmetry, as the revolution axis of the disk and the z axis (beam direction) match in the initial configuration. Thus, considering the temporal evolution of the electric field vector for this polarization, rotating the disk about the y axis, and subsequently about the x axis, has the same effect as rotating it about the x axis and then about the y axis. The obtained average temperature increase is $\Delta T = 212\ K$.

Looking at the Fig. S20c, the orientation response of the rod is studied. As in the case of the disk, a similar symmetry in the temperature increase distribution can be observed. The average temperature increase is remarkably lower than the one obtained in the disk, resulting in a minor tolerance to rotations, hence, showing a worse overall response.

In Fig. S20d, the case of the toroidal particle is analysed. The corresponding colormap exhibits weaker temperature increments for any $\phi$ value upon rotation in $\theta$. Similarly to the disk, it offers a more stable thermal behaviour. Reduction of temperature increments is observed symmetrically for rotations in $\phi$ and $\theta$, achieving a stable value of around 250 K for a full 90º rotation in either $\phi$ or $\theta$. Attending unpolarized nature of the incident light beam, there always exists a configuration in which the electric field vector is parallel to the toroid transversal length, allowing the excitation of the corresponding mode. However, for each time in which the previous situation occurs, the opposite also takes place, with the electric field vector orthogonal to the toroid transversal length, avoiding the possible excitation mode. This leads to a commitment between both configurations, giving rise to a more stable excitation in time. The average temperature increment obtained is $\Delta T = 269$ K, proving that toroidal nanoparticles have a high tolerance to rotations.

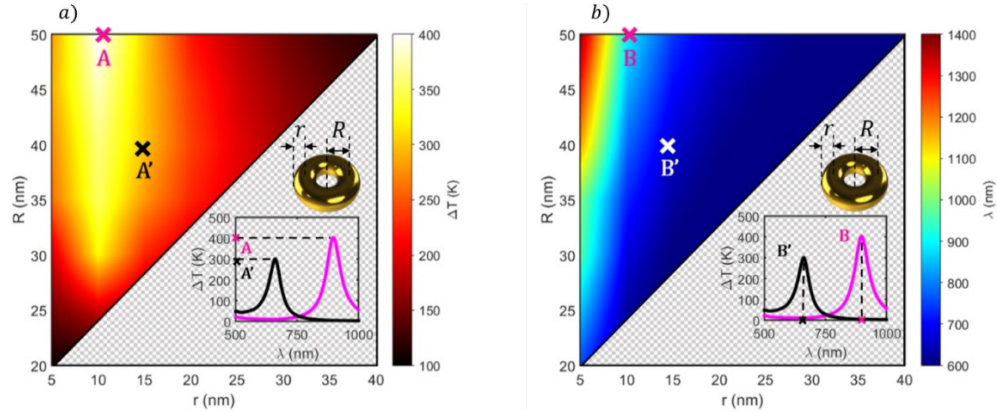

*Figure S21: Thermal response of a full syntonised toroidal nanoparticle in air. a) Maximum temperature increase of toroid for main (R) and secondary (r) radius variations represented between 0 and 400 K. b) Corresponding excitation wavelength of the maximum thermal response as a function of the main and secondary radii. The points A and B correspond to the spectral maximum temperature increase (inset) for the toroid shown in Fig. S19, R=50 nm and r=10 nm. Another example is given by A' and B'.*

Figure S21 displays two colormaps that correspond to the temperature increment produced by the toroid in air (Fig. S21a) and the wavelength at which the optimal resonance occurs for each case (Fig. S21b). Regarding design, main radii lower than 50 nm, more fitting to current nanoparticle sizes, have been considered. In addition, a secondary radius bigger than 5 nm has been assumed, as in the case of Fig. 4.

In sight of Fig. S21a, the tendency that both maximum temperature increase and excitation wavelength exhibit with respect to the structure dimensions can be observed. Note that the optimal temperature is reached when the secondary radius is fixed to 10 nm. On the one hand, larger secondary radii result in geometries closer to a sphere, and so does the temperature response. On the other hand, secondary radii smaller than 10 nm would lead to reduced toroid cross-sections, hindering the support of strong electromagnetic modes that can produce high thermal responses. Looking at the main radius, however, it can be observed that in general the temperature increments grow with it, as the toroid length is increased, which enhances the structure electrical resistance and its corresponding resistive losses.

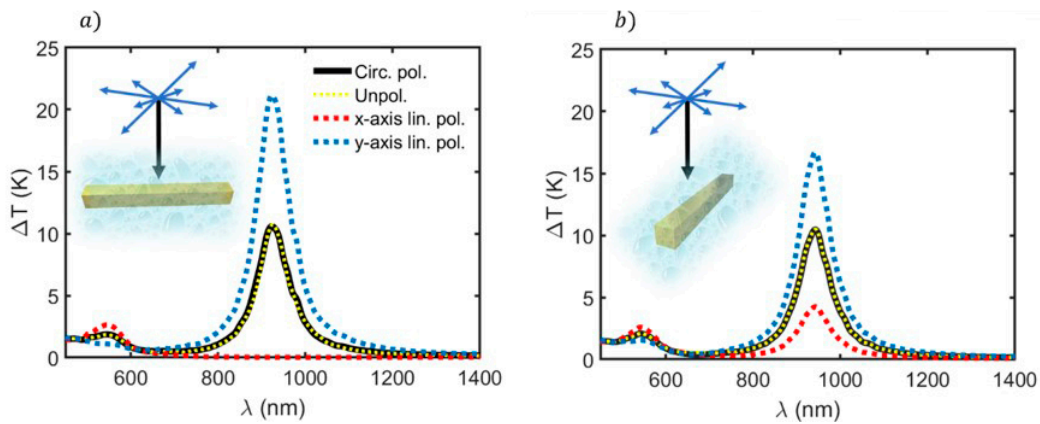

*Figure S22: Thermal response in water of a perfectly oriented rectangular rod (a) and a 30º x-axis and 60º y-axis rotated rod (b) excited by x-polarized (red dotted line), y-polarized (blue dotted line), circular (black solid line) and unpolarized light (yellow dotted line). The dimensions of the rectangular rod are 120 nm length with a section length of 30 nm and for the toroid has a main radius of 50 nm and a second radius of 10 nm.*

Figure S22 justifies the use of circularly polarized light in this work. It is used to emulate the influence that the unpolarized illumination would have on the particle thermal responses. The most common way of simulating unpolarized light consist in performing two different simulations for orthogonal input electric fields, and taking the mean of the results of such simulations. However, executing only one simulation considering circular polarization, results in a higher computational efficiency. As can be seen, the rectangular rod thermal responses are equivalent for circular and unpolarized light, both deriving in a dramatic thermal efficiency reduction with respect to the most favorable y-axis linear polarized light.